# Design of *n*-AlInN on *p*-silicon heterojunction solar cells

Blasco R., Naranjo F. B. and Valdueza-Felip S.

*Abstract*—Aluminum Indium Nitride (AlInN) alloys offer great potential for photovoltaic devices thanks to their wide direct bandgap energy that covers the solar spectrum from 0.7 eV (InN) to 6.2 eV (AlN), and their superior resistance to high temperatures and high-energy particles. In this paper, we report the design of AlInN on silicon heterojunctions, with the aim to explore their potential for solar cell devices through the analysis and optimization of the properties of the AlInN on Si heterojunction. In particular, we study the influence of the AlInN bandgap energy, AlInN thickness and carrier concentration, silicon surface recombination, interface defects and Si wafer quality on the photovoltaic properties (conversion efficiency and external quantum efficiency) of the AlInN on Si heterojunctions. The effect of introducing an anti-reflective coating is also studied. Optimized AlInN on Si heterostructure shows a conversion efficiency of 18% under 1-sun AM1.5G illumination for low-quality Si wafers, which increases to 23.6% for high-quality Si wafers and incorporating a properly designed anti-reflective layer. In comparison with standard Si solar cells without AlInN, the external quantum efficient of the devices increases for wavelengths below 500 nm, making them appealing for space applications. These results lead to the AlInN on Si heterojunction a promising future as a novel technology for solar cell devices.

*Index Terms*—AlInN, silicon, sputtering, theoretical simulations, solar cells, heterojunctions.

## I. Introduction

NOWADAYS, the industrial development and the increase of the world population lead to a rising energy demand year by year. Thus, the limited energy resources and its harvesting is one of the leading and growing industrial sector in our days. In particular, solar energy is available in abundance and can provide the vast energy needs. In this sense, research on solar cells field is focused on achieving low-cost-effective devices, being one of the main objectives to improve devices based on silicon homojunctions.

In that point, III-nitrides are promising materials for developing a new generation of photovoltaic devices using them as active layers in multijunction solar cells [1–3]. Nitride materials offer the advantage of physical and chemical stability leading to high resistance to high temperature and high-energy particle radiation. Besides, their optical properties [4] - like their wide direct bandgap energy tunable from the near infrared (InN: 0.7 eV [5]) to the deep ultraviolet (AlN: 6.2 eV [6]) - are a good advantage to complete devices based on silicon homojunctions, making them able to collect photons in a range of the solar spectrum that silicon cannot do.

Moreover, the use of the radio frequency (RF) sputtering deposition technique [7-9, 30] allows obtaining cheaper and larger devices than other deposition techniques like molecular beam epitaxy (MBE) [10] or metal-organic vapor phase epitaxy (MOCVD) [11]. Besides, this technique allows depositions at low temperature (even at room temperature) with an adequate crystalline quality [12]. However, III-nitride materials deposited by this technique show lower structural quality and higher defects concentration than the ones deposited by others epitaxial techniques. Besides, the presence of impurities like oxygen [13] and defects like nitrogen vacancies [14] induce to a non-intentional doping that leads to a residual carrier concentration close to $10^{21}$ cm-3 causing a Burnstein-Moss effect in the layers [15].

In this sense, it has been proposed In-rich AlInN layers deposited by RF-sputtering to complement the Si photovoltaic properties and to create a low-cost solar cell with a high efficiency [9, 16]. In particular, this group has previously reported studies on AlInN deposited by RF-sputtering on sapphire [17] and Si (111) [18] with a bandgap energy ranging from 1.7 eV (729 nm) to 2.1 eV (590 nm). In these studies, the influence of the substrate temperature and the power applied to the Al target on the structural, morphological and optical properties of $Al_xIn_{1-x}N$ layers ($x \sim$ 0-0.37) is reported. Besides, we have developed devices based on $Al_{0.37}In_{0.63}N$/Si heterojunctions and studied the effect of introducing an AlN

This work was supported by projects Nitpho (TEC2014-60483-R), Anomalos (TEC2015-71127-C2-2-R), NERA (RTI2018-101037-B-I00) the Comunidad de Madrid and FEDER program SINFOTON-2 (S2018/NMT-4326). R. Blasco acknowledges the financial support of his associated to the Ramon y Cajal Fellowship RYC-2013-14084.

R. Blasco and the other authors are with the Photonic Engineering Group of the Electronic Department of the University of Alcalá in Alcalá de Henares, 28871 SPAIN (e-mail: Rodrigo.blasco@uah.es).

buffer on their structural and electrical properties [30], obtaining a device efficiency that evolves from 0.006 to 1.5% depending on the AlN buffer thickness [19]. On the other hand, we also have studied the effect of the AlInN thickness (from 50 to 120 nm) [20] on the photovoltaic properties of $Al_xIn_{1-x}N$ on Si heterojunctions ($x \sim 0.38$-$0.42$) obtaining an efficiency that increases from 1.80 to 2.45%.

In this article, we study the effect of incorporating an AlInN layer to a silicon wafer to improve the efficiency of the Si solar cell. In that sense, we analyze the effect of the AlInN properties as function of the Al content (bandgap energy and background doping), the effect of the AlInN on Si heterojunctions properties (surface recombination and interface defects) and the layer thickness on the photovoltaic properties (conversion efficiency and external quantum efficiency) of the device.

To analyze the effect of those parameters, several simulations were carried out using the Pc1d software [21, 22]. This software allows modeling and understanding all the fundamental effects present in solar cells using Maxwell-Boltzmann (M-B) statistics, becoming one of the most commonly used for photovoltaic devices. However, inasmuch as the Pc1d software does not have nitrides materials, firstly we have introduced their parameters into the software materials library. In order to verify this implementation, we have then compared simulations of InGaN-based junctions with results from the literature.

The optimized AlInN/Si heterojunction, obtained from our simulations, shows the high potential of AlInN layers for Si-based solar cells, obtaining an efficiency of 18% and 23.6% depending on the silicon crystal quality. This efficiency is higher than the one showed by a standard silicon homojunctions of 21.7% for high quality silicon, confirming the improvement when incorporating the AlInN layer to the device. Besides, the high AlInN bandgap energy on top of the lower Si one, results in a significant enhancement of the external quantum efficiency for wavelengths below 500 nm, in comparison with a standard Si solar cell, being promising results for devices specially designed for space applications.

## II. MATERIAL PARAMETERS

Simulations were performed using the Pc1d software [21, 22], a computer program that solves the fully coupled nonlinear equations for the quasi-one-dimensional transport of electrons and holes in crystalline semiconductors devices, with emphasis on photovoltaic devices. Thereby, with this software, we can obtain the photovoltaic characteristics of the devices, namely the current density-voltage (J-V) curve and the external quantum efficiency (EQE). Moreover, the software permits to extract the conduction and valence band alignment at the heterojunction. We needed to include III-nitride semiconductors into the Pc1d software since their properties were not added in the database. For this aim, we first introduced the main optical and electrical properties for the binary compounds (InN, GaN and AlN) and then, we extracted the ones for the ternary compounds depending on the alloy composition. To check the suitability of using the Pc1d software for solar cells including III-nitride layers, we started by simulating InGaN-based homo and heterojunctions. These simulations were compared with the ones obtained by Fabien et al. on the same structures [23], in order to validate the III-nitrides inclusion into the Pc1d library.

## III. CHECKING THE PC1D FOR III-NITRIDES

The main optical and electrical properties of AlN, GaN and InN, used for the simulations, are listed in Table I and II. Concretely, Table I was used to simulate the InGaN-based devices and compare the results with the ones of Fabien *et al.* [23], while Table II was used to simulate the AlInN/Si heterojunctions.

To obtain the parameters of the ternary compounds we followed the Vegard's law. However, to obtain the bandgap energy, the Vegard´s law has been modified introducing a parameter that adequately adjusts the variation of the bandgap energy with the concentration of the ternary compound. This parameter is called bowing parameter (b). The used modified Vegard's law is:

$$a_{AB} = xa_A + (1-x)a_B - bx(1-x) \quad (1)$$

where x is the alloy mole fraction, $a_A$ and $a_B$ are the parameters of the binary compounds and b is the bowing parameter. For the $In_xGa_{1-x}N$ layers, a bowing parameter of 1.43 was selected taking into account the value obtained from Wu *et al.* [24] and Fabien *et al* [23].

The intrinsic carrier concentration of the ternary compounds was calculated using the Melissinos formula:

$$n_i^2 = N_c N_v e^{-\frac{E_g}{K_B T}} \quad (2)$$

where *Nc* and *Nv* are the effective density of states in the conduction and valence band respectively, $K_B$ is the Boltzmann constant, T is the temperature and $E_g$ is the bandgap energy of the alloy. For ternary compounds, the Vegard´s law was applied to obtain $Nc$ and $Nv$. Pc1d software also requires the absorption coefficient of the material as function of the wavelength. In this case, the absorption coefficient for InGaN layers was obtained using the equation:

$$\alpha(E) = \alpha_0 * \sqrt{a(E - E_g) + b(E - E_g)^2} \quad (3)$$

where $\alpha_0 = 2 \times 10^5 \text{cm}^{-1}$ for GaN, the parameter E is the incoming photon energy and the parameters *a* and *b* are fitting parameters obtained from [23].

TABLE I

| Material | GaN | InN |
|---|---|---|
| Bandgap energy (eV) [23] | 3.42 | 0.65 |
| Electron affinity (eV) [23] | 4 | 5.6 |
| Refractive index [39] | 2.3 | 2.9 |
| Effective electron mass ($m_0$) [23] | 0.2 | 0.05 |
| Effective hole mass ($m_0$) [23] | 1.25 | 0.6 |
| Electron mobility ($cm^2$/Vs) [23] | 1000 | 1100 |
| Hole mobility ($cm^2$/Vs) [23] | 170 | 340 |
| Dielectric constant [23] | 8.9 | 10.5 |
| Effective density of states in the conduction band Nc [23] | $3 \times 10^{17}$ | $3 \times 10^{17}$ |
| Effective density of states in the valence band Nv [23] | $2 \times 10^{17}$ | $8 \times 10^{18}$ |
| Intrinsic concentration at 300 K ($cm^{-3}$) | $1.9 \times 10^{-10}$ | $2.3 \times 10^{13}$ |

Table I: Summary of the material parameters of GaN and InN included in the Pc1d software to develop the simulations of InGaN homo and heterojunctions and compare them with the ones of Fabien et al. [23].

The efficiency of the devices under 1 sun of AM 1.5G illumination was estimated from the simulated J-V curves using:

$$E_{ff} = \frac{V_{oc}J_{sc}FF}{P_{in}} \quad (4)$$

where the Fill Factor (FF) is the ratio of maximum obtainable power to the product of the open-circuit voltage ($V_{oc}$) and the short-circuit current ($J_{sc}$) while the power intensity ($P_{in}$) is the input power, which corresponds to one sun spectra of AM 1.5G (1000 W/m$^2$). The EQE is extracted as the ratio between the number of collected carriers and the input optical density (incident photons) as a function of the wavelength.

In order to validate the material parameters of III-nitrides in the program, we have compared our simulation results on InGaN-based homo and heterojunctions, with the ones obtained by Fabien et al. on the same structures [23]. Results are listed in Table III.

In particular, we have simulated two structures based on: 50 nm of *p*-doped In$_{0.25}$Ga$_{0.75}$N ($p \sim 4 \times 10^{18}$ cm$^{-3}$) followed by 100 and 500 nm (S1 and S2 in Table III, respectively) of non-intentionally doped In$_{0.25}$Ga$_{0.75}$N ($n \sim 1 \times 10^{17}$ cm$^{-3}$) and 1 µm of *n*-doped GaN layer ($n \sim 8 \times 10^{18}$ cm$^{-3}$).

From the comparison between our simulations and the ones obtained in [23], we deduce that our results on conversion efficiency are slightly over-estimated by a ~6%, which is in the error bar of the fact of using two different simulation programs with different simulation mesh. In fact, the program used by Fabien et al. [23] is a solver based on 2D/3D finite element analysis of electrical, optical and thermal properties of compound semiconductor devices. The simulator solves self-consistently the Poisson's equation, the current continuity equations, the carrier-energy transport equations, and the quantum mechanical wave equations.

TABLE III

| Structure | Reference | $V_{oc}$ (V) | $J_{sc}$ (mA/cm$^2$) | FF (%) | η (%) |
|---|---|---|---|---|---|
| S1 | This work | 1.87 | 4.36 | 87.08 | 7.1 |
| | Fabien et al. | 1.85 | 4.00 | 90.54 | 6.7 |
| S2 | This wok | 1.89 | 5.85 | 87.14 | 9.6 |
| | Fabien et al. | 1.85 | 5.80 | 86.41 | 9.5 |

Table III: Comparison of the simulation results obtained for different InGaN-based junctions with Pc1d software and the ones obtained by Fabien et al. [23].

*A. AlInN parameters*

Once verified the capability of Pc1d software to simulate heterostructures based on III-nitrides, the extension of the parameters to Al$_x$In$_{1-x}$N layers with $x = 0 - 0.48$ was carried out taking into account the values for the binary compounds of Table II. The method of obtaining the parameters is identical to the one explained in the previous section, except for the absorption coefficient, the bandgap energy and the carrier concentration, which was obtained from experimental measurements in samples deposited by RF-sputtering.

Moreover, a minority carrier lifetime of ~1 ps was used for the AlInN in all simulations. This value was introduced assuming a lifetime of 30 ps for InN [25]. The value for the alloy is lower in order to take into account the alloy disorder and the high density of impurities and defects present in layers deposited by RF-sputtering.

The effect of the surface recombination of the AlInN is also analyzed in this study. In that sense, the experimental values of III-nitrides compounds deposited by MBE are around 10$^3$ cm/s [26]. However, inasmuch as this study is based on ternary compounds deposited by sputtering with a lower structural quality, the value assigned to the simulation is 10$^8$ cm/s.

The bandgap energy Eg, was estimated from transmission measurements of AlInN samples deposited on sapphire by RF-sputtering with Al mole fraction in the range $x = 0$ to 0.48, as obtained from x-ray diffraction measurements [27], as illustrated in Fig. 1(a).

Concretely, $E_g$ is obtained through a linear fit of the squared absorption coefficient as a function of the photon energy [dashed line of Fig. 1(a)]. The absorption of the samples was estimated from the transmission spectra considering the relationship $\alpha(E) \propto -\ln(Tr)$, which neglects optical scattering and reflection losses. From this method, a variation of the bandgap energy, obtained from the transmittance measurements of samples grown on sapphire [see inset of Fig. 1(a)], varies from $E_g = 1.7$ eV (708 nm) for InN to $E_g = 2.6$ eV (476 nm) for $Al_xIn_{1-x}N$ ($x \approx 0.48$). As it is explained before, the high values of $E_g$ are related to band filling and a blue shift of the bandgap energy of the alloy (Burstein-Moss effect) caused by an unintentional doping caused by impurities like oxygen [13]. The obtained values for $E_g$ are close (with an error below 2%) to the calculated one using Eq. 2 using a bowing parameter of 5.1, which has been experimentally obtained for $Al_xIn_{1-x}N$ layers deposited by RF-sputtering [15].

TABLE II

| Material | AlN | InN |
|---|---|---|
| Bandgap energy (eV) [39] | 6.2 | 1.7** |
| Electron affinity (eV) [39] | 1.9 | 5.6 |
| Refractive index [39] | 2.15 | 2.9 |
| Effective electron mass (m$_0$) [39] | 0.3 | 0.07 |
| Effective hole mass (m$_0$) [39] | 3.5 | 1.63 |
| Electron mobility (cm$^2$/Vs) [39] | 135 | 8.7** |
| Hole mobility (cm$^2$/Vs) [39] | 14 | 340* |
| Dielectric constant [39] | 8.5 | 15.3 |
| Effective density of states in the conduction band Nc [39] | 6.3×10$^{18}$ | 9×10$^{17}$ |
| Effective density of states in the valence band Nv [39] | 4.8×10$^{20}$ | 5.3×10$^{19}$ |
| Intrinsic concentration at 300 K (cm$^{-3}$) | 9.4×10$^{-34}$ | 2.3×10$^{13}$ |

Table II: Summary of the material parameters of AlN and InN included in the Pc1d software to develop the simulations of the AlInN/Si heterojunctions. Notes: *Obtained from Fabien et al. [23]. **Obtained from experimental measurements.

Furthermore, in order to obtain realistic values from the simulations, some parameters of the AlInN ternary compounds were obtained from experimental data. In particular, the Pc1d software requires the absorption coefficient and carrier concentration of the layers and its mobility. These parameters were obtained from transmission and Hall Effect measurements of AlInN layers grown on sapphire by RF-sputtering for each Al mole fraction.

In the case of the carrier mobility, Hall Effect measurements point to a decrease of the electron mobility from 8.7 to 0.1 $cm^2/Vs$ when the Al content increases from 0 to 48%. On the other hand, the mobility of the holes is assumed to be smaller due to its higher effective mass as (see Table II).

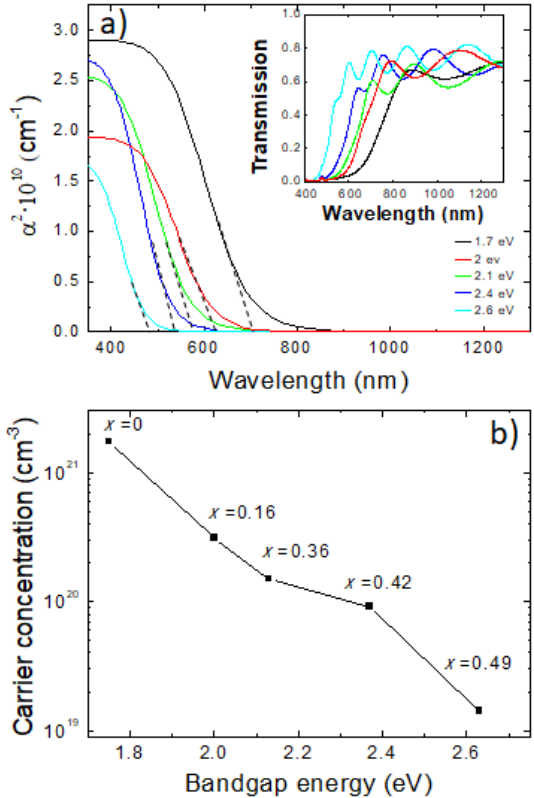

Fig. 1. (a) Absorption coefficient and transmittance (inset) spectra of the AlInN on sapphire samples as a function of the AlInN bandgap energy. (b) Carrier concentration vs the bandgap energy for the same samples. The Al mole fraction of each AlInN layer is also marked.

## IV. SIMULATION RESULTS AND DISCUSSION

The simulated main structure consists of an *n*-type $Al_xIn_{1-x}N$ layer ($x = 0 - 0.48$) on a *p*-Si substrate. As a start, we have chosen an AlInN thickness of 300 nm and a carrier concentration density that evolves in the range of $10^{19} - 10^{21}$ $cm^{-3}$ decreasing with the Al mole fraction as shown in Fig. 1(b). As is explained before, a conservative AlInN recombination time of 1 ps and a high surface recombination rate of 108 cm/s were chosen due to the properties of the alloy deposited by RF-sputtering.

As *p*-type Si substrate, we have considered a commercial 500-μm thick wafer with a resistivity of 10–100 Ω·cm, as provided by the manufacturer datasheet (Semiconductor Wafer INC company). A surface recombination rate of $10^8$ cm/s and a recombination time of 200 μs have been considered in order to simulate a low-quality silicon wafer [28,29].

The effect of the recombination of the silicon surface will be studied throughout this work. For this purpose, the simulations start with a value of $10^8$ cm/s, a quite higher value than the usually assigned to a low silicon quality $10^6$ cm/s [30].

On the other hand, it has been reported that a highly doped channel can be formed during the deposition of III-nitrides on silicon due to diffusion of metallic atoms inside the substrate during growth [31, 32]. To account for this channel, an interface defect density of $10^7$ $cm^{-3}$/peak following an error function (ERC) at the first 7 nm of the silicon substrate was introduced to obtain a more realistic model.

Starting from this structure, we have analyzed the effect of the following parameters: AlInN bandgap energy, thickness and background doping; interface defects and Si surface recombination rate; and anti-reflective coating, on the photovoltaic properties of the heterojunction. The effect of each one is separately studied while the rest parameters are kept fixed. The optimized structure will be compared with a Si-based *p-n* homojunction, where the *p*-silicon layer is the same than the *p*-silicon of the optimized AlInN/Si heterojunction, and the *n*-silicon layer has the same thickness and carrier concentration than the AlInN layer, in order to evaluate its performance improvement. The area of the devices was fixed to 1 $cm^2$ for all the simulations.

### A. Effect of AlInN bandgap energy

The bandgap energy and the transparency of the AlInN material is of crucial importance in the design of the heterojunction in order to maximize the number of photons that will be absorbed by each layer and the number of generated photocarriers that will be collected as a function of the wavelength.

As said previously, the AlInN bandgap energy was obtained through transmission measurements of $Al_xIn_{1-x}N$ on sapphire samples deposited with Al mole fraction in the range $x = 0$ to 0.48, as obtained from x-ray diffraction measurements [27] [see

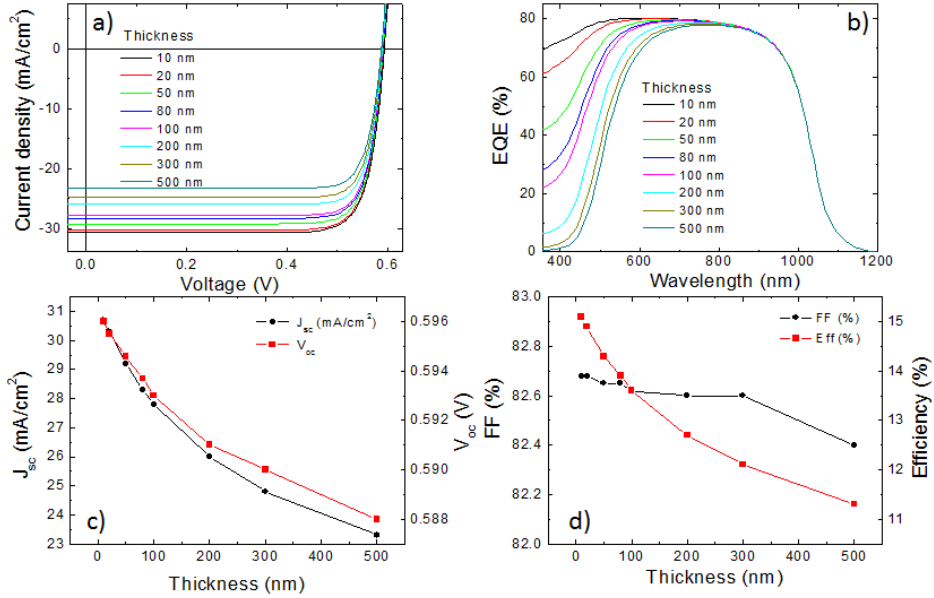

Fig. 3. J-V (a) and EQE (b) curves of the AlInN on Si heterojunctions as function of the AlInN thickness. The evolution of the $V_{oc}$ and $J_{sc}$, FF and efficiency vs the AlInN thickness is plotted in (c) and (d), respectively.

Fig. 1(a) and 1(b)]. The data of carrier concentration were obtained through Hall Effect measurements from AlInN on sapphire samples deposited by RF-sputtering [17], obtaining a decrease of the n-type carrier concentration from $2\times10^{21}$ to $2\times10^{19}$ cm$^{-3}$ accordingly to the bandgap energy, as illustrated in Fig. 1(b).

The simulated current density vs voltage (J-V) curves under 1 sun of AM 1.5G illumination and the EQE spectra of the samples are presented in Fig. 2 (a) and (b), respectively. The increase of the Al bandgap energy by 0.85 eV results in an increase of $V_{oc}$ from 0.24 to 0.59 V, and of $J_{sc}$ from 8.8 to 24.8 mA/cm$^2$ [Fig. 2(c)]. This enhancement of $J_{sc}$ is accompanied by a blue shift of the EQE cutoff at short wavelengths according to the change of absorption band-edge energy of the AlInN layer with the Al content. At the same time, an enhancement of the maximum EQE from 48% at 925 nm to 78% at 750 nm for n-InN/p-Si and n-Al$_{0.48}$In$_{0.52}$N/p-Si devices is also observed. This behavior leads to a FF and conversion efficiency improvement from 69 to 82.7% and from 1.8 to 12.1%, respectively [Fig. 2(d)]. Taking into account these results, the Al mole fraction was kept fixed for the rest of the study to $x = 0.48$ ($E_g = 2.6$ eV), with a carrier concentration $n \sim 2\times10^{19}$ cm$^{-3}$. It should be pointed out that the maximum Al content in the layers is limited to $x = 0.48$ due to the insulator character showed by non-intentionally doped AlInN layers with $x > 0.48$ deposited using RF-sputtering by our group [33].

### B. Effect of the AlInN thickness

The photovoltaic properties of the AlInN on Si heterojunctions were further analyzed by varying the AlInN thickness from 500 nm to 10 nm, while maintaining the rest of parameters fixed. Figures 3 (a) and (b) show the illuminated J-V curves under 1 sun of AM 1.5G illumination and the EQE spectra of the samples, respectively. The value of $V_{oc}$ and the FF remain almost constant around 0.59 V and 82.5%, respectively; while $J_{sc}$ rises from 23.3 to 30.4 mA/cm$^2$ when reducing the AlInN thickness from 500 to 10 nm, leading to an increase of the conversion efficiency from 11.3 to 15.2% [Fig. 3 (c) y (d)]. This effect is correlated with an increase of the EQE at short wavelengths when decreasing the AlInN thickness, which is attributed to an increase of photon absorption at AlInN layer close to the depletion region of the heterojunction. On the opposite, for higher AlInN thickness, the absorption close the surface but far from the depletion region disables the electron-hole pair separation by the internal field and its further collection at the contacts. From now on, we will fix the AlInN thickness to 10 nm.

### C. Effect of the AlInN n-type doping

In the following study, the carrier concentration was changed from $n = 1\times10^{17}$ to $1\times10^{21}$ cm$^{-3}$ considering a 10-nm thick AlInN on Si heterojunction with $E_g = 2.6$ eV. The simulated J-V curves [shown in Fig. 4 (a)] point to a constant $J_{sc}$ for the whole analyzed range of carrier concentration, whereas the $V_{oc}$ shows an increase from 0.34 to 0.60 V up to $n = 1\times10^{19}$ cm$^{-3}$, as expected from the increase of the built-in voltage of the heterojunction with the donor concentration. In our case, a built-in voltage of ~ 0.24 V was obtained using the following equation:

$$V_{built-in} = \frac{K_B T}{q} \ln\left(\frac{N_D N_A}{n_i^2}\right) \quad (5)$$

where $N_D$ and $N_A$ are the donor (n-type) and acceptor (p-type) concentration, respectively, $n_i$ is the intrinsic carrier concentration, $K_B$ is the Boltzmann constant, q is the electron charge and T is the temperature.

The simulation fails for carrier concentration above this number as the Pc1d software uses the M-B statistics, which does not describe properly the case of degenerate semiconductors. This case can be approximated by an adequate increase in the bandgap energy of the degenerate semiconductor, as it has been done implicit in previous sections considering the experimental $E_g$ values of samples with Burstein-Moss shift. In the case of the study of this section, for carrier concentration above $n = 1\times10^{19}$ cm$^{-3}$, a change of $E_g$ should be introduced to obtain realistic simulations.

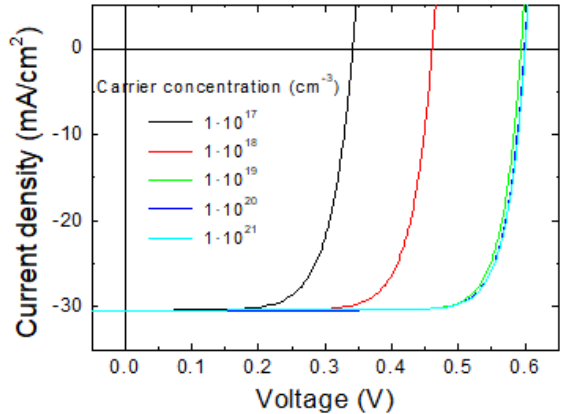

Fig. 4. J-V (a) curves of the AlInN on Si heterojunctions as function of the AlInN background doping.

The obtained increase of the $V_{oc}$, for carriers concentration below $n = 1\times10^{19}$ cm$^{-3}$ leads to an increase of the FF from 70 to 83% and thus, of the conversion efficiency of the heterojunction from 7.3% for a carrier concentration of $n = 1\times10^{17}$ cm$^{-3}$ to 15.2% for a carrier concentration of $n = 1\times10^{19}$ cm$^{-3}$. The EQE spectra present similar shape and values independently of the AlInN doping level.

Although the maximum efficiency is obtained for high doping levels of $n = 1\times10^{21}$ cm$^{-3}$ ($\eta$ = 15.3%), the fact that the program does not describe rightly degenerate semiconductors leads us to not consider this value. Therefore, and given that the efficiency variation is not significant, it has been decided to maintain a carrier concentration of $n = 10^{19}$ cm$^{-3}$ that corresponds to a bandgap energy approximated to the maximum obtained in the previous study of $E_g$ = 2.6 eV [Fig. 1(b)].

### D. Effect of interface defects

A thin p+ doped layer at the first Si nanometres can emerge due to the Al diffusion during our AlInN deposition process. Some studies show an interface diffusion of GaN and AlN above $10^{17}$ cm$^{-3}$ [34,35]. However, in order to understand the effect of this thin layer at the junction interface we have studied the device performance for a range of defects density between 0 and $10^{19}$ cm$^{-3}$/peak.

Figures 5 (a) and (b) show the illuminated J-V curves and the EQE spectra of the samples under study as a function of the interface defects. In this figure, it can be observed that the EQE drops in the 600 to 1000 nm spectral range for a density of defects range above $10^{18}$ cm$^{-3}$, reducing the $V_{oc}$ from 0.60 to 0.27 V, the $J_{sc}$ from 30.4 to 16.9 mA/cm$^2$ and accordingly the conversion efficiency from 2.62 to 15.2% [Fig. 5(d)].

We point out that the field generated by p+ zone at the interface hinders the electron drift towards the *n*-type semiconductor, reducing the EQE and the efficiency of the heterojunction. From now on, with the aim to obtain an optimized structure, we select a structure without interface defects.

### E. Effect of surface recombination rate

One important issue in the development of Si-based devices is the passivation of the Si surface to reduce the effect of the Si dangling bonds. High-efficient devices present a reduced carrier recombination at the Si surface, thus increasing the carrier lifetime and the conversion efficiency [36].

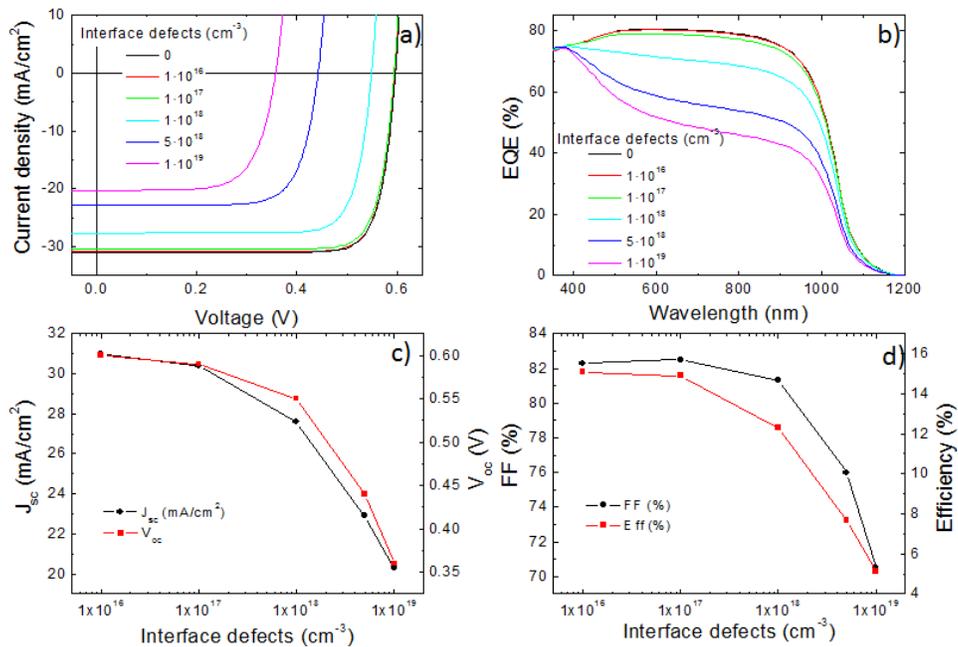

Fig. 5. J-V (a) and EQE (b) curves of the AlInN on Si heterojunctions as function of the interface defects. The evolution of the $V_{oc}$ and $J_{sc}$, FF and efficiency vs the interface defects is plotted in (c) and (d), respectively.

To study the effect of the Si surface passivation on AlInN on Si heterojunctions, we have modified the surface recombination rate of both sides of the silicon layer from S = $10^8$ cm/s (low-quality Si) to S = 10 cm/s (high-quality Si) [27]. Figures 6 (a) and (b) show the J-V and EQE curves obtained in this analysis.

In particular, an enhancement of the $V_{oc}$ from 0.54 to 0.67 V and of $J_{sc}$ from 30.7 to 32.2 mA/cm$^2$ is shown when decreasing the Si surface recombination. On the other hand, a reduced surface recombination results on a moderate improvement of the EQE at short wavelengths, and in the 900-1000 nm wavelength range, near the Si cut-off. Those improvements lead to an increase of conversion efficiency from 13.6 to 18% when reducing S from $10^8$ to 10 cm/s. This effect is due to the improved electron collection in both faces of silicon because of the reduction of the dangling bond density [37]. From now on, we will fix the Si surface recombination of the 10-nm thick Al$_{0.48}$In$_{0.52}$N on Si heterojunctions to S = 10 cm/s.

*F. Effect of surface recombination rate*

*G. Effect of the silicon wafer quality*

Considering previous results, a maximum conversion efficiency of the devices of 18% under 1 sun AM1.5G illumination has been obtained. However, as it was noted at the beginning of section 3, a Si with a low crystal quality was firstly selected. Here, we study the effect of increasing the quality of the Si substrate by adjusting the values of resistivity and bulk recombination time [27]. Concretely, the Si resistivity was varied from 100 to 1 Ω·cm and the bulk recombination time from 200 to 2000 µs [28]. Meanwhile, the Si surface recombination was maintained to S = 10 cm/s, assuming a properly passivated surface, taking into account the results presented in section 3.5.

Table IV shows the electrical characteristics of the optimized AlInN on Si heterojunctions as a function of the Si wafer quality. We can observe an enhancement of $V_{oc}$ from 0.67 to 0.71 V and of $J_{sc}$ from 32.2 to 32.7 mA/cm$^2$ while maintaining the FF of the J-V curve, leading to a conversion efficiency to 19.7%.

*H. Effect of anti-reflective layer*

The conversion efficiency of the AlInN on Si heterojunctions can be further improved reducing the optical losses of the device caused by a high light reflectance at the top surface. The introduction of an anti-reflective layer (ARL) at the surface reduces the light reflectance at a certain wavelength range, improving the transmission of the light to the structure.

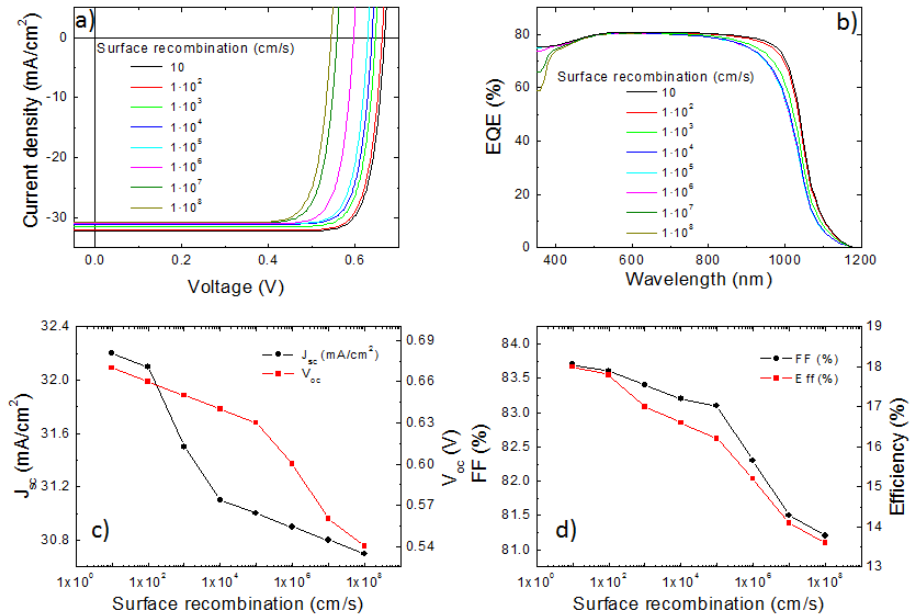

Fig. 6. J-V (a) and EQE (b) curves of the AlInN on Si heterojunctions as function of the surface recombination. The evolution of the $V_{oc}$ and $J_{sc}$, FF and efficiency vs the surface recombination is plotted in (c) and (d), respectively.

The transmission of light at a certain wavelength range is modulated by its refractive index and thickness. The ARL refractive index, $n_2$, depends on the refractive index of the materials of each side, namely air ($n_0 = 1$) and $Al_{0.48}In_{0.52}N$ ($n_1 = 2.54$) following $n_2 = \sqrt{n_0 n_1}$ [38]. The refractive index of the $Al_{0.48}In_{0.52}N$ layer was estimated following Vergard´s and assuming that $n_{InN} = 2.9$ and $n_{AlN} = 2.15$ [39]. The ARL thickness is one quarter the wavelength of the incoming wave and depends on $n_2$ like $d = \frac{\lambda_0}{4n_2}$ [38].

Taking into account that the maximum radiation of the solar spectrum corresponds to the wavelength region of 450 to 700 nm, the ARL is designed to improve the EQE around 600 nm, obtaining a refractive index of $n_2 = 1.59$ and a thickness of d = 94 nm. . Further optimization of these parameters with simulations using Pc1d considering the optimized device leads to the final design of ARL with a refractive index of 1.6 and a thickness of 100 nm.

Figure 7(a) shows the EQE vs the wavelength of the optimized AlInN on Si heterojunction with high quality Si with and without the designed ARL. The incorporation of the ARL reduces the surface reflectance of the device in the 600-nm wavelength range, increasing the peak EQE value from 81 to 99.6%. This improvement leads to an increase of the $J_{sc}$ from 32.7 to 38.9 mA/cm$^2$ and of the device efficiency reaching a value of 23.6% (see Table IV).

This efficiency is close to the intrinsic limit of the silicon solar cells ~29% [40] and to others technologies such perovskites ~27% [41] or complex InGaN/Si tandem structures, with a conversion efficiency of ~31% [3], leading to the AlInN on Si heterojunction a promising future as novel technology for solar cell devices.

## V. COMPARISON OF THE OPTIMIZED ALINN/SI HETEROJUNCTION WITH A P-N SOLAR CELL

The previously optimized AlInN on Si heterojunction, which optimized parameters are listed in the inset of Fig. 7(a), is compared with a Si homojunction solar cell to analyze the improvements achieved by the use of a III-nitride layer in a photovoltaic Si-based device. For this comparison, the Si homojunction is based on a p-n Si structure with the same high-quality p-Si than the one used for the optimized n-AlInN/p-Si structure, and a high-quality n-Si layer with the same thickness and carrier concentration then for the AlInN layer (10 nm and 1×10$^{19}$ cm$^{-3}$, respectively).

Figure 7 (b) shows the comparison between the EQE of the optimized AlInN on Si heterojunction and the Si homojunction, both with high quality Si and ARL. We observe that the EQE of the AlInN on Si heterojunction presents a higher EQE in the wavelength region below 500 nm. This region is important because it covers the peak of the AM1.5G and AM1.0 solar spectra, as shown in the plot. This gain in EQE leads to a $J_{sc}$ increase from 34.9 to 38.9 mA/cm$^2$ while the $V_{oc}$ remains almost constant. These trends result in an increase in conversion efficiency from 21.7 to 23.6% (see Table IV).

Besides, this enhancement is especially effective for devices operating outside the atmosphere with AM1.0 illumination, pointing that AlInN/Si heterojunction are a promising novel technology for space applications.

## VI. CONCLUSIONS

Here we explored the potential of n-AlInN on p-silicon heterojunctions for solar cell devices through the design and optimization of the structure. We studied the influence of the AlInN bandgap energy, thickness and carrier concentration, Si surface recombination, interface defects, Si wafer quality and anti-reflective coating on the photovoltaic properties of the devices. Best results were obtained for a 10-nm thick AlInN on Si junction with an AlInN bandgap energy of 2.63 eV, an n-type doping of 10$^{19}$ cm$^{-3}$, a surface recombination of 10 cm/s and no interface defects. Optimized AlInN on Si heterostructures show a conversion efficiency of 18% under 1-sun AM1.5G illumination for low-quality Si wafers, which increases to 23.6% for high-quality Si wafers and incorporating a properly designed anti-reflective layer. In comparison with standard Si solar cells without AlInN, the external quantum efficient of the devices increases for wavelengths below 500 nm, making them interesting for space applications.

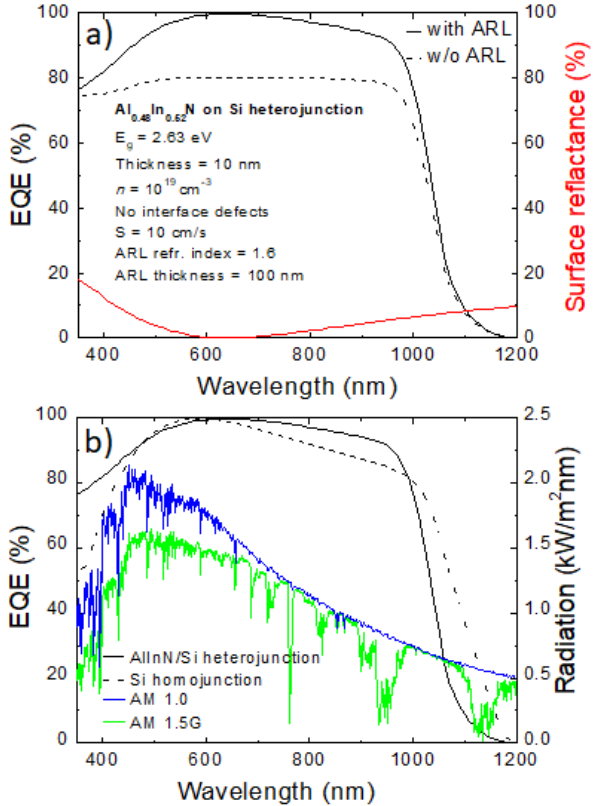

Fig. 7. a) EQE spectra of the optimized AlInN on Si heterojunctions with and without ARL. Surface reflectance of the device with ARL. Inset: parameters of the optimized AlInN on Si structure. b) Comparison of the EQE between the optimized AlInN on Si heterojunction and the Silicon homojunction, both with high quality Si and ARL. The AM1.5G an AM1.0 sun radiation spectra are also plotted.